\documentclass[prl,twocolumn,superscriptaddress]{revtex4-2}
\pdfoutput=1

\usepackage{graphicx}
\usepackage{subfigure}
\usepackage{dcolumn}
\usepackage{bm}
\usepackage{bbm}
\usepackage{amsthm}
\usepackage{thmtools}

\usepackage{mathtools}
\usepackage{physics}
\usepackage{binarytree}
\usepackage{tikz,pgfplots}
\usepackage{verbatim}
\usepackage{pgfplots}
\usepackage{placeins}
\usepackage{algorithm2e}
\usetikzlibrary{decorations.pathreplacing}
\usetikzlibrary{arrows.meta}
\usetikzlibrary{graphs}
\usepackage{xcolor}
\usepackage{rotating}
\usepackage{yquant}
\usepackage{tabularray}
\usepackage[colorlinks=true, citecolor=red, urlcolor=red]{hyperref}
\RestyleAlgo{ruled}
\pgfplotsset{compat=1.16}

\begin{document}

\title[The optimized Quantum Autoencoder] {Optimized Quantum Autoencoder}

\author{Yibin Huang}
\affiliation{Institute of Physics, Beijing National Laboratory for Condensed
  Matter Physics,\\Chinese Academy of Sciences, Beijing 100190, China}
\affiliation{School of Physical Sciences, University of Chinese Academy of
  Sciences, Beijing 100049, China}

\author{Muchun Yang}
\affiliation{Institute of Physics, Beijing National Laboratory for Condensed
  Matter Physics,\\Chinese Academy of Sciences, Beijing 100190, China}
\affiliation{School of Physical Sciences, University of Chinese Academy of
  Sciences, Beijing 100049, China}

\author{D. L. Zhou} \email[]{zhoudl72@iphy.ac.cn}
\affiliation{Institute of Physics, Beijing National Laboratory for Condensed
  Matter Physics,\\Chinese Academy of Sciences, Beijing 100190, China}
\affiliation{School of Physical Sciences, University of Chinese Academy of
  Sciences, Beijing 100049, China}

\date{\today}

\begin{abstract}
   Quantum autoencoder (QAE) compresses a bipartite quantum state into its subsystem by a self-checking mechanism. How to characterize the lost information in this process is essential to understand the compression mechanism of QAE\@. Here we investigate how to decrease the lost information in QAE for any input mixed state. We theoretically show that the lost information is the quantum mutual information between the remaining subsystem and the ignorant one, and the encoding unitary transformation is designed to minimize this mutual information.  Further more, we show that the optimized unitary transformation can be decomposed as the product of a permutation unitary transformation and a disentanglement unitary transformation, and the permutation unitary transformation can be searched by a regular Young tableau algorithm. Finally we numerically identify that our compression scheme outperforms the quantum variational circuit based QAE\@.
\end{abstract}                         
\maketitle

\textit{Introduction.---}
Information compression is fundamental in classical and quantum information processing~\cite{nielsen2010quantum,Wilde_2017,https://doi.org/10.1002/j.1538-7305.1948.tb01338.x,PhysRevLett.93.230504,PhysRevA.51.2738,doi:10.1080/09500349414552191,PhysRevA.66.022311}.
As an effective tool for classical information compression, the classical autoencoders have wide applications in feature extraction \cite{7965877, 10.1145/1390156.1390294}, denoising \cite{persee.fr:intel_0769-4113_1987_num_2_1_1804, 10.5555/1756006.1953039}, and data compression \cite{10.1007/BF00332918, Hinton1993AutoencodersMD}. Quantum autoencoder (QAE), as its extension from the classical regime to the quantum regime, was proposed and successfully demonstrated by Romero et al\@. for compressing an ensemble of pure states~\cite{Romero_2017}. The goal of QAE is to compress quantum data from $d_{A}\times d_{B}$-dimensional Hilbert space to $d_{B}$-dimensional Hilbert space using a unitary transformation $U$, which is selected by approximately recovering the quantum data from the compressed data through a decoder. QAEs have been implemented experimentally using photons \cite{PhysRevLett.122.060501,PhysRevA.102.032412} and superconducting qubits \cite{https://doi.org/10.1002/qute.201800065}.

Recently QAE has served as a useful tool for lots of applications, such as quantum error correction(QEC)\cite{Locher2023quantumerror,PhysRevA.103.L040403,PhysRevLett.124.130502, PhysRevA.103.L040403}, quantum phase transitions \cite{PhysRevE.107.045301,PhysRevE.97.013306,PhysRevA.108.063303},
information scrambling\cite{PhysRevB.101.064406,PhysRevResearch.3.L032057}, anomaly detection \cite{PhysRevB.108.165408,PhysRevResearch.3.043184,PhysRevD.105.095004}, qubit readout\cite{PhysRevApplied.20.014045}, complexity of quantum states \cite{PhysRevB.106.L041110} and detecting families of quantum many-body
scars (QMBS)\cite{PhysRevB.105.224205}. In addition, Ref. \cite{PhysRevResearch.5.023039} investigates QAE using entropy production. Ref.~\cite{Patel_2023} examines information loss, resource costs, and run time from practical application of QAE\@.

As QAE utilizes variational quantum circuits for optimization, there is considerable interest in its experimental performance, with few conducting theoretical analyses \cite{PhysRevResearch.5.023039,PhysRevApplied.15.054012} of QAE. To fill this gap, in this Letter, we theoretically analyze the process of compressing quantum data using QAE. The main focus of this paper is to investigate the following questions: $(i)$ What information is lost during the compression process? $(ii)$ Theoretically, what is the optimal compression strategy? 

In this Letter, we theoretically show that the lost information is the quantum mutual information between the remaining subsystem and the ignorant one.
The compression process of QAE effectively reduces the mutual information between two parts. At the same time, we theoretically provide the optimal compression strategy for reducing mutual information.
We prove that there exists an optimal unitary transformation which can be decomposed as the product of a permutation unitary transformation and a disentanglement unitary transformation. The permutation unitary transformation can be searched by a regular Young tableau algorithm which is superior to the variational quantum circuit algorithm.

\begin{figure}[htbp]
\vspace{3mm}
\centering
\includegraphics[width=0.45\textwidth]{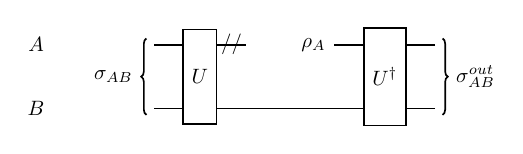}
\caption{In this model, the quantum state $\sigma_{AB}$ undergoes unitary evolution U, followed by the removal of subsystem A to obtain the compressed quantum state $\sigma_{B}$. By taking the tensor product with the quantum state $\rho_{A}$ on subsystem A and then applying the inverse evolution of the original unitary transformation, the quantum state is restored, we ultimately obtain the quantum state $\sigma^{\text{out}}_{AB}$.
}
\label{fig:0}
\end{figure}

\textit{Mixed-State Compression Model.---}
In quantum information, information source can be described by a mixed state. Based on the QAE, we propose a compression model for a quantum mixed state (quantum information source), as illustrated in Fig.~\ref{fig:0}.
In this model, a bipartite quantum state $\sigma_{AB}$ undergoes a unitary evolution $U$, the state becomes   $\sigma^{U}_{AB} = U\sigma_{AB}U^{\dagger}$,
 whose reduced states for the subsystem $A$ and the subsystem $B$ are given respectively by
$\sigma^{U}_{A}  = \Tr_{B}\sigma^{U}_{AB},\sigma^{U}_{B}  = \Tr_{A}\sigma^{U}_{AB}.$
 Then the state of subsystem $A$ is removed, and the state of subsystem $B$ is $\sigma^U_{B}$. To recover the initial state $\sigma_{AB}$, an auxiliary quantum state $\rho_{A}$ on subsystem $A$ is introduced and the inverse evolution of the original unitary transformation is applied,  and the ultimate quantum state
$\sigma^{\text{out}}_{AB}=U^{\dagger}[\rho_{A}\otimes \sigma^U_B]U.$
 Our aim is to find a unitary evolution $U$ and a density matrix $\rho_{A}$ to minimize the quantum relativity between $\sigma_{AB}$ and $\sigma^{\text{out}}_{AB}$, i.e.,
\begin{equation}
  \label{eq:1}
  \min_{U,\rho_A} S(\sigma_{AB}||\sigma^{\text{out}}_{AB}).
\end{equation}
 Here we use the quantum relative entropy instead of the fidelity to measure the difference between two quantum states.
On one hand, compression of the quantum information source is inherently discussed in the ensemble sense, and functions related to entropy can better characterize it. On the other hand, relative entropy can measure the difference between two states. The statistical interpretation of quantum relative entropy is: it tells us how difficult it is to distinguish the state $\rho$ from $\sigma$ \cite{hiai1991proper}. 
In fact, the quantum relative entropy (and its generalizations) can induce a Riemannian metric on the space of quantum states, making it a mathematical Riemannian space \cite{10.1063/1.533053}. 

This is why we use quantum relative entropy to measure the difference between $\rho$ and $\sigma$.
Quantum relative entropy can also be used to quantify the amount of resource in a
resource theory \cite{PhysRevLett.121.190503, PhysRevLett.121.190504}, to study uncertainty relation \cite{PhysRevE.109.L012103, PhysRevA.102.022217} and to quantify the non-Gaussian character of a quantum state \cite{PhysRevA.78.060303}.

\textit{Minimization over the auxiliary state.---}
\label{sec:optim-over-auxil}
We divide the minimization~\eqref{eq:1} into two steps. In the first step, we will minimization over the auxiliary state $\rho_A$ for any fixed unitary transformation $U$. In the second step, we will minimization over the unitary transformation $U$ for the optimized auxiliary state obtained in the first step. In this section, we will complete the first step by giving the main result in the following theorem.

\textit{Theorem 1.---}
 For any unitary transformation $U$,
 \begin{equation}
   \label{eq:2}
   \min_{\rho_A} S(\sigma_{AB}||\sigma^{\mathrm{out}}_{AB}) = S^U(A:B),
 \end{equation}
 where $S^U(A:B)$ is the quantum mutual entropy of $\sigma^U_{AB}$, defined by
$S^{U}(A:B)=S(\sigma^{U}_{A})+S(\sigma^{U}_{B})-S(\sigma^{U}_{AB}).$
The minimization is taken if and only if
$\rho_{A}=\sigma^{U}_{A}.$

\textit{Proof.---}
We prove it by the following evaluation:
\begin{align*}
        & \phantom{=}  S(\sigma_{AB}\Vert\sigma^{\text{out}}_{AB})\\
        &=S(U\sigma_{AB}U^{\dagger}\Vert U\sigma^{\text{out}}_{AB}U^{\dagger})\\
        &=S(\sigma^{U}_{AB}\Vert(\rho_{A}\otimes\sigma^{U}_{B}))\\
        &=\Tr(\sigma^{U}_{AB}\log\sigma^{U}_{AB})-\Tr[\sigma^{U}_{AB}\log(\rho_{A}\otimes\sigma^{U}_{B}))]\\
        &=\Tr(\sigma^{U}_{AB}\log\sigma^{U}_{AB})-\Tr[\sigma^{U}_{AB}(\log\rho_{A}+\log\sigma^{U}_{B})]\\
        &=\Tr(\sigma^{U}_{AB}\log\sigma^{U}_{AB})-\Tr_{A}\sigma^{U}_{A}\log\rho_{A}-\Tr_{B}\sigma^{U}_{B}\log\sigma^{U}_{B}\\
        &\geq \Tr(\sigma^{U}_{AB}\log\sigma^{U}_{AB})-\Tr_{A}\sigma^{U}_{A}\log\sigma^{U}_{A}-\Tr_{B}\sigma^{U}_{B}\log\sigma^{U}_{B}\\
        &=S(\sigma^{U}_{A})+S(\sigma^{U}_{B})-S(\sigma^{U}_{AB})\\
        &=S^{U}(A:B).
\end{align*}
In the seventh line of the above calculation, we utilize the Klein's inequality. According to the Klein's equality, we obtain that the equality is satisfied if and only if $\rho_A=\sigma^U_A$.
Therefore, for a given $U$, the minimum value of $S(\sigma_{AB}\Vert\sigma^{\text{out}}_{AB})$ is $S^{U}(A:B)$, and the condition for achieving this minimum is given by $\rho_A=\sigma^U_A$.

Following Eq.~\eqref{eq:2}, the minimization in Eq.~\eqref{eq:1} is equivalent to the minimization
\begin{equation}
    \label{eq:3}
    \min_{U,\rho_A} S(\sigma_{AB}||\sigma^{\text{out}}_{AB}) = \min_{U}S^{U}(A:B). 
\end{equation}
Eq.~\eqref{eq:3} implies that for a given unitary transformation $U$, the minimal difference (measured by the quantum relative entropy) between the input state $\sigma_{AB}$ and the output state $\sigma^{\text{out}}_{AB}$ equals to the bipartite correlation (measured by the mutual entropy) in the middle state $\sigma^U_{AB}$.

\textit{Transforming unitary transformation into permutation.---}
In this section, we will come to the second step of our minimization: find a unitary transformation $U$ to minimize the mutual entropy $S^U(A:B)$. To minimize the mutual entropy, we adopt the following strategy: first find a unitary transformation $V_D$ to disentangle the state $\sigma_{AB}$ into a separable state $\sigma^{V_D}_{AB}$; second find a second unitary $V_{\tau}$ to minimize the classical correlation in the separable state $\sigma^{V_D}_{AB}$. Note that the similar strategy has been used in the study of the distribution of mutual entropy in a tripartite pure state by Yang, et al.~\cite{PhysRevA.108.052402}. Here we will show the strategy works exactly in our case, which is explicitly given by the following theorem.

Before state our theorem, we introduce two relative concepts: a disentanglement unitary transformation and a permutation unitary transformation. To define a disentanglement unitary transformation, we start with the eigen decomposition of $\sigma_{AB}$, $\sigma_{AB}=\sum^{d_{A}d_{B}}_{\alpha=1}p_{\alpha}\ket{\phi_{\alpha}}\bra{\phi_{\alpha}},$
  where $d_A$ ($d_B$) is the dimension of the Hilbert space of subsystem $A$ ($B$), $p_{\alpha}$ and $\ket{\phi_{\alpha}}$ are eigen value and eigenvector of the state $\sigma_{AB}$ respectively. Then we make a one-one map from $\alpha \to i,m$, with $i \in \{ 1,2,\cdots,d_{A} \}$ and $m \in \{ 1,2,\cdots,d_{B} \}$, and define a disentanglement unitary transformation $V_{D}\ket{\phi_{\alpha}}=\ket{i}\otimes\ket{m}=\ket{im},$
which transforms $\sigma_{AB}$ into a separable state $\sigma^{V_{D}}_{AB}=V_{D}\sigma_{AB}V^{\dagger}_{D}=\sum_{im}p_{im}\ket{im}\bra{im}.$
Next we define a permutation unitary transformation
$V_{\tau}\ket{im}=\ket{\tau(im)},$
where $\tau$ is a permutation operation for the set $\{i m\}$.
Now we are ready to give our main theorem.

\textit{Theorem 2.---}
\label{thm_2}
The minimization of mutual entropy by a unitary transformation
\begin{equation}
\label{min}
    \min_{U}S^{U}(A:B)=\min_{V_{\tau}}S^{V_{\tau}V_{D}}(A:B),
\end{equation}
where $V_{\tau}$ is a permutation unitary transformation, and $V_D$ is a disentanglement unitary transformation.

\textit{Proof.---}
If we denote $V=UV^{-1}_{D}$, then $U=VV_{D}$. Eq.~\eqref{min} can be written in a more direct form
$\min_{V}S^{V V_D}(A:B)=\min_{V_{\tau}}S^{V_{\tau}V_{D}}(A:B).$
Because the von Neumman entropy is invariant under any unitary transformation, i.e., $S(\sigma^{V V_D}_{AB})=S(\sigma^{V_{\tau}V_{D}}_{AB})=S(\sigma_{AB})$,  then we only need to prove
\begin{equation}
\label{th1_uu}
\min_{V}S(\sigma^{V V_D}_{A})+S(\sigma^{V V_D}_{B})=\min_{V_{\tau}}S(\sigma^{V_{\tau}V_{D}}_{A})+S(\sigma^{V_{\tau}V_{D}}_{B}).
\end{equation}
The state $\sigma_{AB}^U$ can be written in the form
$\sigma_{AB}^U = \sigma_{AB}^{V V_D} = \sum_{i m} p_{i m} V|i m\rangle \langle i m|V^{\dagger}.$
Assuming that its reduced state $\sigma^{U}_{A}$ is diagonal in the basis $\{ \ket{i^{U}},~1\leq i\leq d_{A} \}$. and $\sigma^{U}_{B}$ is diagonal in the basis $\{ \ket{m^{U}},~1\leq m\leq d_{B} \}$, i.e.,
$\sigma^{U}_{A}=\sum^{d_{A}}_{j=1}p^{U}_{Aj}\ket{j^{U}}\bra{j^{U}},~\sigma^{U}_{B}=\sum^{d_{B}}_{n=1}p^{U}_{Bn}\ket{n^{U}}\bra{n^{U}},$
where
$p^U_{A j} = \sum_{i m n} p_{i m} p^{U}(j n|i m),~p^U_{B n} = \sum_{i j m} p_{i m} p^{U}(j n|i m),$
with
$p^{U}(jn|im) = \bra{j^{U}n^{U}}V\ket{im}\bra{im}V^{\dagger}\ket{j^{U}n^{U}}.$
Note that $p^{U}(jn|im)$ is a doubly stochastic matrix. Especially when $U=V_{\tau}V_D$, we can take $|j^U n^U\rangle$ to be $|j n\rangle$, and
\begin{equation}
  \label{eq:6}
  p^{V_{\tau} V_D}(jn|im)=
  \begin{cases}
    1 & \text{ if } jn = \tau(im), \\
    0 & \text{ otherwise}.
  \end{cases}
\end{equation}
In other words, $p^{V_{\tau} V_D}$ is a permutation matrix.  According to the Birkhoff's theorem~\cite{matrix}, the matrix $p^U(jn|im)$ can be written as
$p^{U}(jn|im) = \sum_{\tau} \lambda^{U}_{\tau}p^{V_{\tau} V_D}(jn|im),$
where $\lambda^U_{\tau}\ge 0$ and $\sum_{\tau} \lambda^U_{\tau}=1$. Hence
$p^U_{A j} = \sum_{\tau} \lambda^U_{\tau} p^{V_{\tau} V_D}_{A j},~p^U_{B n} = \sum_{\tau} \lambda^U_{\tau} p^{V_{\tau} V_D}_{B n},$
which implies that
$\sigma^U_A = \sum_{\tau} \lambda^U_{\tau} W^U_A \sigma^{V_{\tau} V_D}_A W^{U\;\dagger}_A, ~\sigma^U_B = \sum_{\tau} \lambda^U_{\tau} W^U_B \sigma^{V_{\tau} V_D}_B W^{U\;\dagger}_B.$
where $W^U_A|i\rangle = |i^U\rangle$ and $W^U_B|m\rangle = |m^U\rangle$, which are local unitary transformations. According to the concavity of the von Neumman entropy~\cite{nielsen2010quantum}, we get
\begin{align}
  \label{eq:7}
  S(\sigma^U_{A}) & \ge \sum_{\tau} \lambda^U_{\tau} S(W^U_A \sigma^{V_{\tau} V_D}_A W^{U\;\dagger}_A) = \sum_{\tau} \lambda^U_{\tau} S(\sigma^{V_{\tau} V_D}_A), \\
  S(\sigma^U_{B}) & \ge \sum_{\tau} \lambda^U_{\tau} S(W^U_B \sigma^{V_{\tau} V_D}_B W^{U\;\dagger}_B) = \sum_{\tau} \lambda^U_{\tau} S(\sigma^{V_{\tau} V_D}_B),
\end{align}
which implies that
\begin{equation}
    \sum_{\tau} \lambda^{U}_{\tau}(S(\sigma^{V_{\tau} V_D}_A)+S(\sigma^{V_{\tau} V_D}_B))\leq S(\sigma^{U}_{A})+S(\sigma^{U}_{B}).
\end{equation}
Therefore for any unitary transformation $U$, there must exist a unitary transformation $V_{\tau}$ such that
$S(\sigma^{V_{\tau} V_D}_A)+S(\sigma^{V_{\tau} V_D}_B)\leq S(\sigma^{U}_{A})+S(\sigma^{U}_{B})$,
which implies that the minimization (\ref{th1_uu}) can be realized by some $V_{\tau}$.

Theorem 2 transforms our problem into finding a permutation $\tau$ that minimizes $S^{V_{\tau}V_{D}}(A:B)$. In the next section, we will introduce the method for finding such a permutation $\tau$.

\textit{Regular Young Tableau Traversal Algorithm.---}
In this section, we will introduce the procedure for finding the permutation $\tau$. According to Eq.~\eqref{eq:6}, we obtain
\begin{equation}
  \label{eq:10}
  p^{\tau}_{j n} \equiv p^{V_{\tau} V_D}_{j n} = \sum_{i m} p_{i m} p^{V_{\tau} V_D}(j n|i m) = p_{\tau^{-1}(j n)},
\end{equation}
which implies that the unitary transformation $V_\tau$ maps $p_{\tau^{-1}(j n)}$ to $p^{\tau}_{jn}$. In general, permutations $\tau$ and matrices $p^{\tau}$ have a one-to-one correspondence, and the size of the searching space is $(d_{A}d_{B})!$.

Before presenting a theorem to decrease the size of the searching space, we first define the concept of a decreasing matrix: A matrix $M$ is said to be a decreasing matrix if and only if  its element $M_{ij}\geq M_{ab}$ for all $i\leq a$ and $j\leq b$. In particular, a matrix is a decreasing matrix if and only if all the row vectors and all the column vectors in a decreasing matrix are decreasing vectors.

\textit{Theorem 3.---}
\label{thm3}
    For any matrix $p^{\tau}$, we can always find a permutation $\tau^{\prime}$ such that $p^{\tau^{\prime} \circ \tau }$ is a decreasing matrix and
    \begin{equation}
      \label{eq:11}
      S^{V_{\tau} V_D}(A:B) \ge S^{V_{\tau^{\prime} \circ \tau} V_D}(A:B).
    \end{equation}

\textit{Proof.---}
   Let us introduce two basic matrix dependent permutations $\tau_A$ and $\tau_B$: for any matrix $M$, $\tau_A$ ($\tau_B$) permutates the row (column) indexes while keeping the column (row) indexes invariant such that all the row (column) vectors in $M^{\tau_A}$ ($M^{\tau_B}$) become decreasing vectors. We will show the corresponding unitary transformations decrease the mutual entropy:
   \begin{align}
     \label{eq:12}
     S^{V_{\tau} V_D}(A:B) & \ge S^{V_{\tau_A \circ \tau} V_D}(A:B), \\
      S^{V_{\tau} V_D}(A:B) & \ge S^{V_{\tau_B \circ \tau} V_D}(A:B). \label{eq:13}
   \end{align}
   Note that $p^{\tau_A \circ \tau}_{j n} = p^{\tau}_{\tau_A^{-1}(j n)}$. Because $\tau_A$ does not change the column index $n$, we have $p_{B n}^{\tau_A \circ \tau}=p_{B n}^{\tau}$, which implies that $S(\sigma_B^{V_{\tau_A \circ \tau} V_D})=S(\sigma_B^{V_{\tau}\circ V_D})$. For any given $n$, $p^{\tau_A \circ \tau}_{j n}$ is a decreasing vector, which implies that $p_A^{\tau} \prec p_A^{\tau_A \circ \tau}$, where $x \prec y$ is defined by $\sum_{j=1}^k x_j^{\downarrow}\le\sum_{j=1}^k y_{j=1}^k$ for $1\le k\le n$, and when $k=n$, the equality is taken. Because the entropy is Shur-concave, we obtain $S(\sigma_A^{V_{\tau_A\circ \tau} V_D}) \le S(\sigma_A^{V_{\tau} V_D})$. Together with the fact that the entropy of the whole system is invariant under any unitary transformation, we complete our proof of Eq.~\eqref{eq:12}. Similarly, we can prove Eq.~\eqref{eq:13}.

   We repeat the operation $n$ times: $\tau^{\prime}={(\tau_B \circ \tau_A)}^{n}$ until the matrix $p^{\tau^{\prime}\circ\tau}$ becomes a decreasing matrix, which is indeed the fixed point for such operations. This completes our proof.

A direct consequence of Theorem 3 implies that the searching space is composed by all the permutations whose $p^{\tau}$ is a decreasing matrix. Note that $p^{\tau}$ is a rearrangement of $\{p_{\alpha}\}$. Without losing of generality, we assume $p_1\ge p_2\ge \cdots \ge p_{d_A d_B}$, which makes us to build a definite map from the lower index $\alpha$ to  its value $p_{\alpha}$. So we only need partition the index numbers into the $d_A\times d_B$ lattice, which forms a Young tableau. Theorem 3 implies that the searching space can be restricted to the permutations corresponding to regular Young tableau, i.e., the Young tableau whose row vectors and column vectors are increasing vectors.

\begin{table}[htbp]
  \centering
  \begin{tblr}{ccccc}
    \hline
     $d_{A}$ & $2$ & $3$ & $4$ & $5$&\\
    \hline
     $d_{Bmax}$ & $18$ & $9$ & $6$ & $5$& \\
    \hline
    $Y_{num}$ & $5\times 10^{8}$ & $4\times 10^{8}$ & $1\times 10^{8}$ & $3\times 10^{8}$ &\\
    \hline
  \end{tblr}
  \caption{We choose the order of magnitude of $10^{8}$ as a standard, which can be computed using personal computers. With $d_{A}$ fixed, within the computational power range, the maximum value of $d_{B}$ is $d_{Bmax}$. The corresponding numbers of regular Young Tableaux is $Y_{num}$.} \label{tab:numbers}
\end{table}

The number of regular Young tableaux $Y_{\text{num}}$ is given by the Hook Length Formula~\cite{andrews_1984}, which is numerically demonstrated in Table~\ref{tab:numbers}. We choose $(d_A, d_{B \text{max}})$ such that the number $Y_{\text{num}}$ is restricted in the order of $10^8$, which is the maximal number our personal computer can handle by the exhaustive method. Within this range of Hilbert space dimensions, for a given density matrix, we can utilize Algorithm 1 in the supplementary materials \cite{SM} to seek the optimal unitary transformation $V_{\tau}$ to compress the quantum state.

\textit{Search algorithms to reduce mutual information.---}
The number of regular Young tableaux grows exponentially as the dimension of the Hilbert space increases. This renders the traversal algorithm ineffective when the dimension $d_{A}$ and $d_{B}$ exceeds the computational limits.
In this case, although traversing all regular Young tableaux to minimize mutual information is not feasible, we have found an alternative search algorithm to reduce mutual information. 
Our search algorithm consists of two steps. 

The first step is breadth-first search: randomly generate $N_1$ regular Young tableaux, calculate the mutual information corresponding to each tableau, and select the $N_2$ tableaux with the lowest mutual information as the results of our breadth-first search. This algorithm has been written in Algorithm 2 of the supplementary materials \cite{SM}. 

The second step of our algorithm is the depth-first search: starting from a regular Young tableau, we search for its neighboring regular Young tableaux. Among these neighboring tableaux, we find the one with the minimum corresponding mutual information. We repeat this process $N_D$ times to ultimately reduce the mutual information. This algorithm has been written in Algorithm 3 of the supplementary materials \cite{SM}. The feasibility of the depth-first search algorithm stems from the fact that regular Young Tableaux near each other tend to have similar mutual information values. In fact, we have grouped all regular Young Tableaux into different sets based on their proximity. As the dimension of the Hilbert space increases, the number of such sets grows exponentially. Therefore, the breadth-first search process essentially finds tableaux distributed across different sets, while the depth-first search aims to identify the minimum mutual information within these sets.

\begin{figure}
\subfigbottomskip=10pt
\hspace{-4mm}
\subfigure[]{
\includegraphics[width=0.245\textwidth]{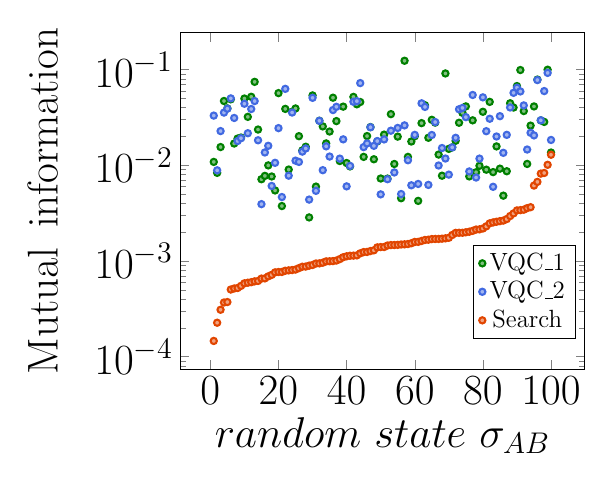}}
\label{2_c_1}
\hspace{-2mm}
\subfigure[]{
\includegraphics[width=0.238\textwidth]{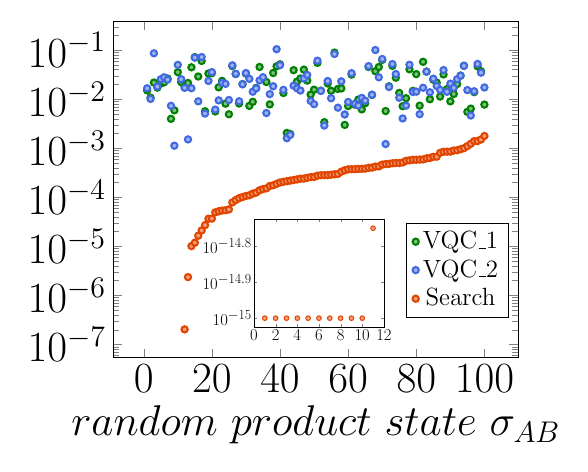}}
\label{2_c_2}
\caption{In our numerical demonstration, we set $d_{A}=d_{B}=8,~N_1=20000,~N_2=12,~N_D=200$. (a) We randomly generate 100 diagonal mixed states on the AB system and optimize their mutual information using two variational quantum circuits (VQC\_1, VQC\_2) and our search algorithm, respectively. Among those mixed states, our algorithm consistently outperforms the other two algorithms. (b) We randomly generated 100 product states, theoretically, their mutual information can be reduced to 0. We utilized these three methods for optimization. On average, our search algorithm reduced the mutual information to 0.00036, while the other algorithms averaged only 0.02371 and 0.02435.}
\label{2_c}
\end{figure}
To demonstrate the superiority of the search algorithm, considering the case where $d_{A}=d_{B}=8$, we compared it with optimization algorithms for variational quantum circuits. We selected the VQC\_1 model and the VQC\_2 model. The structures of two quantum circuits can be found in Sec. \MakeUppercase{\romannumeral3} of the
 Supplemental Material \cite{SM}. In the search algorithm, we set $N_1=20000,~N_2=12,~N_D=200$. The optimization time for the three algorithms is similar. The results are presented in Fig.~\ref{2_c}. In Fig.~\ref{2_c}(a), we randomly generated 100 diagonalized mixed states and optimized them using VQC\_1 and VQC\_2, with the loss function being the mutual information between subsystems A and B. In the quantum variational circuit optimization method, we perform three optimizations for each quantum state, and the final optimization result is the best one among the three optimizations. Subsequently, we used the search algorithm to reduce the mutual information by searching regular Young tableaux, as depicted in Fig.~\ref{2_c}(a). In 100 optimization runs, our algorithm consistently outperformed the other two algorithms.

Finally, we randomly generated 100 diagonalized product states, after rearranging their diagonal elements, we optimized the mutual information between subsystems A and B using VQC\_1 and VQC\_2. Theoretically, the mutual information of these product states can be reduced to 0. The final results are shown in Fig.~\ref{2_c}(b). In the cases where the mutual information is less than $10^{-15}$, we set the optimization result to be $10^{-15}$. On average, our algorithm reduced the mutual information to 0.00036, while the other algorithms averaged only 0.02371 and 0.02435. It is evident that our algorithm is superior.

\textit{Summary.---}
In this Letter, we theoretically show that the lost information in QAE is the quantum mutual information between the remaining subsystem and the ignorant one, and prove that there exists an optimal unitary transformation which can be decomposed into the product of disentanglement and permutation. For any input quantum mixed state, once we fix the disentanglement unitary transformation, our task becomes searching for a permutation in the search space which consists of $(d_{A}d_{B})!$ permutations. Furthermore, we establish a one-to-one correspondence between each permutation and a $d_{A}\times d_{B}$ Young tableau and prove that it is sufficient to search for the optimal permutation only within the set of permutations corresponding to regular Young tableaux.

In practical applications, we categorize the dimensions of subsystems into two cases based on computational capabilities: exhaustible and non-exhaustible. For exhaustible cases, we can use Algorithm 1 of the supplementary materials \cite{SM} to generate all regular Young tableaux in order to search for the optimal permutation. For non-exhaustible cases, we can use the search algorithm of the supplementary materials \cite{SM} to search for relatively optimal permutations. Finally, through numerical comparisons, we find that our algorithm outperforms the variational quantum circuit algorithm.

We expect that our complete informative picture on QAE will increase the understandings on the physical mechanism of QAE, and further broaden its applications in quantum information processing.

\textit{Acknowledgments.---}This work is supported by National Key Research and Development Program of China (Grant No. 2021YFA0718302 and No. 2021YFA1402104),  and National Natural Science Foundation of China (Grants No. 12075310).

\bibliographystyle{apsrev4-2}
\bibliography{bib.bib}

\begin{thebibliography}{45}%
\makeatletter
\providecommand \@ifxundefined [1]{%
 \@ifx{#1\undefined}
}%
\providecommand \@ifnum [1]{%
 \ifnum #1\expandafter \@firstoftwo
 \else \expandafter \@secondoftwo
 \fi
}%
\providecommand \@ifx [1]{%
 \ifx #1\expandafter \@firstoftwo
 \else \expandafter \@secondoftwo
 \fi
}%
\providecommand \natexlab [1]{#1}%
\providecommand \enquote  [1]{``#1''}%
\providecommand \bibnamefont  [1]{#1}%
\providecommand \bibfnamefont [1]{#1}%
\providecommand \citenamefont [1]{#1}%
\providecommand \href@noop [0]{\@secondoftwo}%
\providecommand \href [0]{\begingroup \@sanitize@url \@href}%
\providecommand \@href[1]{\@@startlink{#1}\@@href}%
\providecommand \@@href[1]{\endgroup#1\@@endlink}%
\providecommand \@sanitize@url [0]{\catcode `\\12\catcode `\$12\catcode `\&12\catcode `\#12\catcode `\^12\catcode `\_12\catcode `\%12\relax}%
\providecommand \@@startlink[1]{}%
\providecommand \@@endlink[0]{}%
\providecommand \url  [0]{\begingroup\@sanitize@url \@url }%
\providecommand \@url [1]{\endgroup\@href {#1}{\urlprefix }}%
\providecommand \urlprefix  [0]{URL }%
\providecommand \Eprint [0]{\href }%
\providecommand \doibase [0]{https://doi.org/}%
\providecommand \selectlanguage [0]{\@gobble}%
\providecommand \bibinfo  [0]{\@secondoftwo}%
\providecommand \bibfield  [0]{\@secondoftwo}%
\providecommand \translation [1]{[#1]}%
\providecommand \BibitemOpen [0]{}%
\providecommand \bibitemStop [0]{}%
\providecommand \bibitemNoStop [0]{.\EOS\space}%
\providecommand \EOS [0]{\spacefactor3000\relax}%
\providecommand \BibitemShut  [1]{\csname bibitem#1\endcsname}%
\let\auto@bib@innerbib\@empty
\bibitem [{\citenamefont {Nielsen}\ and\ \citenamefont {Chuang}(2010)}]{nielsen2010quantum}%
  \BibitemOpen
  \bibfield  {author} {\bibinfo {author} {\bibfnamefont {M.}~\bibnamefont {Nielsen}}\ and\ \bibinfo {author} {\bibfnamefont {I.}~\bibnamefont {Chuang}},\ }\href@noop {} {\emph {\bibinfo {title} {Quantum Computation and Quantum Information}}}\ (\bibinfo  {publisher} {Cambridge university press},\ \bibinfo {year} {2010})\BibitemShut {NoStop}%
\bibitem [{\citenamefont {Wilde}(2017)}]{Wilde_2017}%
  \BibitemOpen
  \bibfield  {author} {\bibinfo {author} {\bibfnamefont {M.~M.}\ \bibnamefont {Wilde}},\ }\href@noop {} {\emph {\bibinfo {title} {Quantum Information Theory}}},\ \bibinfo {edition} {2nd}\ ed.\ (\bibinfo  {publisher} {Cambridge University Press},\ \bibinfo {year} {2017})\BibitemShut {NoStop}%
\bibitem [{\citenamefont {Shannon}(1948)}]{https://doi.org/10.1002/j.1538-7305.1948.tb01338.x}%
  \BibitemOpen
  \bibfield  {author} {\bibinfo {author} {\bibfnamefont {C.~E.}\ \bibnamefont {Shannon}},\ }\href {https://doi.org/https://doi.org/10.1002/j.1538-7305.1948.tb01338.x} {\bibfield  {journal} {\bibinfo  {journal} {Bell System Technical Journal}\ }\textbf {\bibinfo {volume} {27}},\ \bibinfo {pages} {379} (\bibinfo {year} {1948})},\ \Eprint {https://arxiv.org/abs/https://onlinelibrary.wiley.com/doi/pdf/10.1002/j.1538-7305.1948.tb01338.x} {https://onlinelibrary.wiley.com/doi/pdf/10.1002/j.1538-7305.1948.tb01338.x} \BibitemShut {NoStop}%
\bibitem [{\citenamefont {Devetak}\ \emph {et~al.}(2004)\citenamefont {Devetak}, \citenamefont {Harrow},\ and\ \citenamefont {Winter}}]{PhysRevLett.93.230504}%
  \BibitemOpen
  \bibfield  {author} {\bibinfo {author} {\bibfnamefont {I.}~\bibnamefont {Devetak}}, \bibinfo {author} {\bibfnamefont {A.~W.}\ \bibnamefont {Harrow}},\ and\ \bibinfo {author} {\bibfnamefont {A.}~\bibnamefont {Winter}},\ }\href {https://doi.org/10.1103/PhysRevLett.93.230504} {\bibfield  {journal} {\bibinfo  {journal} {Phys. Rev. Lett.}\ }\textbf {\bibinfo {volume} {93}},\ \bibinfo {pages} {230504} (\bibinfo {year} {2004})}\BibitemShut {NoStop}%
\bibitem [{\citenamefont {Schumacher}(1995)}]{PhysRevA.51.2738}%
  \BibitemOpen
  \bibfield  {author} {\bibinfo {author} {\bibfnamefont {B.}~\bibnamefont {Schumacher}},\ }\href {https://doi.org/10.1103/PhysRevA.51.2738} {\bibfield  {journal} {\bibinfo  {journal} {Phys. Rev. A}\ }\textbf {\bibinfo {volume} {51}},\ \bibinfo {pages} {2738} (\bibinfo {year} {1995})}\BibitemShut {NoStop}%
\bibitem [{\citenamefont {Jozsa}\ and\ \citenamefont {Schumacher}(1994)}]{doi:10.1080/09500349414552191}%
  \BibitemOpen
  \bibfield  {author} {\bibinfo {author} {\bibfnamefont {R.}~\bibnamefont {Jozsa}}\ and\ \bibinfo {author} {\bibfnamefont {B.}~\bibnamefont {Schumacher}},\ }\href {https://doi.org/10.1080/09500349414552191} {\bibfield  {journal} {\bibinfo  {journal} {Journal of Modern Optics}\ }\textbf {\bibinfo {volume} {41}},\ \bibinfo {pages} {2343} (\bibinfo {year} {1994})},\ \Eprint {https://arxiv.org/abs/https://doi.org/10.1080/09500349414552191} {https://doi.org/10.1080/09500349414552191} \BibitemShut {NoStop}%
\bibitem [{\citenamefont {Hayashi}\ and\ \citenamefont {Matsumoto}(2002)}]{PhysRevA.66.022311}%
  \BibitemOpen
  \bibfield  {author} {\bibinfo {author} {\bibfnamefont {M.}~\bibnamefont {Hayashi}}\ and\ \bibinfo {author} {\bibfnamefont {K.}~\bibnamefont {Matsumoto}},\ }\href {https://doi.org/10.1103/PhysRevA.66.022311} {\bibfield  {journal} {\bibinfo  {journal} {Phys. Rev. A}\ }\textbf {\bibinfo {volume} {66}},\ \bibinfo {pages} {022311} (\bibinfo {year} {2002})}\BibitemShut {NoStop}%
\bibitem [{\citenamefont {Meng}\ \emph {et~al.}(2017)\citenamefont {Meng}, \citenamefont {Catchpoole}, \citenamefont {Skillicom},\ and\ \citenamefont {Kennedy}}]{7965877}%
  \BibitemOpen
  \bibfield  {author} {\bibinfo {author} {\bibfnamefont {Q.}~\bibnamefont {Meng}}, \bibinfo {author} {\bibfnamefont {D.}~\bibnamefont {Catchpoole}}, \bibinfo {author} {\bibfnamefont {D.}~\bibnamefont {Skillicom}},\ and\ \bibinfo {author} {\bibfnamefont {P.~J.}\ \bibnamefont {Kennedy}},\ }in\ \href {https://doi.org/10.1109/IJCNN.2017.7965877} {\emph {\bibinfo {booktitle} {2017 International Joint Conference on Neural Networks (IJCNN)}}}\ (\bibinfo {year} {2017})\ pp.\ \bibinfo {pages} {364--371}\BibitemShut {NoStop}%
\bibitem [{\citenamefont {Vincent}\ \emph {et~al.}(2008)\citenamefont {Vincent}, \citenamefont {Larochelle}, \citenamefont {Bengio},\ and\ \citenamefont {Manzagol}}]{10.1145/1390156.1390294}%
  \BibitemOpen
  \bibfield  {author} {\bibinfo {author} {\bibfnamefont {P.}~\bibnamefont {Vincent}}, \bibinfo {author} {\bibfnamefont {H.}~\bibnamefont {Larochelle}}, \bibinfo {author} {\bibfnamefont {Y.}~\bibnamefont {Bengio}},\ and\ \bibinfo {author} {\bibfnamefont {P.-A.}\ \bibnamefont {Manzagol}},\ }in\ \href {https://doi.org/10.1145/1390156.1390294} {\emph {\bibinfo {booktitle} {Proceedings of the 25th International Conference on Machine Learning}}},\ \bibinfo {series and number} {ICML '08}\ (\bibinfo  {publisher} {Association for Computing Machinery},\ \bibinfo {address} {New York, NY, USA},\ \bibinfo {year} {2008})\ p.\ \bibinfo {pages} {1096–1103}\BibitemShut {NoStop}%
\bibitem [{\citenamefont {Fogelman-Soulié}\ and\ \citenamefont {Le~Cun}(1987)}]{persee.fr:intel_0769-4113_1987_num_2_1_1804}%
  \BibitemOpen
  \bibfield  {author} {\bibinfo {author} {\bibfnamefont {F.}~\bibnamefont {Fogelman-Soulié}}\ and\ \bibinfo {author} {\bibfnamefont {Y.}~\bibnamefont {Le~Cun}},\ }\href {https://doi.org/10.3406/intel.1987.1804} {\bibfield  {journal} {\bibinfo  {journal} {Intellectica}\ }\textbf {\bibinfo {volume} {2}},\ \bibinfo {pages} {114} (\bibinfo {year} {1987})},\ \bibinfo {note} {included in a thematic issue : Apprentissage et machine}\BibitemShut {NoStop}%
\bibitem [{\citenamefont {Vincent}\ \emph {et~al.}(2010)\citenamefont {Vincent}, \citenamefont {Larochelle}, \citenamefont {Lajoie}, \citenamefont {Bengio},\ and\ \citenamefont {Manzagol}}]{10.5555/1756006.1953039}%
  \BibitemOpen
  \bibfield  {author} {\bibinfo {author} {\bibfnamefont {P.}~\bibnamefont {Vincent}}, \bibinfo {author} {\bibfnamefont {H.}~\bibnamefont {Larochelle}}, \bibinfo {author} {\bibfnamefont {I.}~\bibnamefont {Lajoie}}, \bibinfo {author} {\bibfnamefont {Y.}~\bibnamefont {Bengio}},\ and\ \bibinfo {author} {\bibfnamefont {P.-A.}\ \bibnamefont {Manzagol}},\ }\href@noop {} {\bibfield  {journal} {\bibinfo  {journal} {J. Mach. Learn. Res.}\ }\textbf {\bibinfo {volume} {11}},\ \bibinfo {pages} {3371–3408} (\bibinfo {year} {2010})}\BibitemShut {NoStop}%
\bibitem [{\citenamefont {Bourlard}\ and\ \citenamefont {Kamp}(1988)}]{10.1007/BF00332918}%
  \BibitemOpen
  \bibfield  {author} {\bibinfo {author} {\bibfnamefont {H.}~\bibnamefont {Bourlard}}\ and\ \bibinfo {author} {\bibfnamefont {Y.}~\bibnamefont {Kamp}},\ }\href {https://doi.org/10.1007/BF00332918} {\bibfield  {journal} {\bibinfo  {journal} {Biol. Cybern.}\ }\textbf {\bibinfo {volume} {59}},\ \bibinfo {pages} {291–294} (\bibinfo {year} {1988})}\BibitemShut {NoStop}%
\bibitem [{\citenamefont {Hinton}\ and\ \citenamefont {Zemel}(1993)}]{Hinton1993AutoencodersMD}%
  \BibitemOpen
  \bibfield  {author} {\bibinfo {author} {\bibfnamefont {G.~E.}\ \bibnamefont {Hinton}}\ and\ \bibinfo {author} {\bibfnamefont {R.~S.}\ \bibnamefont {Zemel}},\ }in\ \href {https://api.semanticscholar.org/CorpusID:2445072} {\emph {\bibinfo {booktitle} {Neural Information Processing Systems}}}\ (\bibinfo {year} {1993})\BibitemShut {NoStop}%
\bibitem [{\citenamefont {Romero}\ \emph {et~al.}(2017)\citenamefont {Romero}, \citenamefont {Olson},\ and\ \citenamefont {Aspuru-Guzik}}]{Romero_2017}%
  \BibitemOpen
  \bibfield  {author} {\bibinfo {author} {\bibfnamefont {J.}~\bibnamefont {Romero}}, \bibinfo {author} {\bibfnamefont {J.~P.}\ \bibnamefont {Olson}},\ and\ \bibinfo {author} {\bibfnamefont {A.}~\bibnamefont {Aspuru-Guzik}},\ }\href {https://doi.org/10.1088/2058-9565/aa8072} {\bibfield  {journal} {\bibinfo  {journal} {Quantum Science and Technology}\ }\textbf {\bibinfo {volume} {2}},\ \bibinfo {pages} {045001} (\bibinfo {year} {2017})}\BibitemShut {NoStop}%
\bibitem [{\citenamefont {Pepper}\ \emph {et~al.}(2019)\citenamefont {Pepper}, \citenamefont {Tischler},\ and\ \citenamefont {Pryde}}]{PhysRevLett.122.060501}%
  \BibitemOpen
  \bibfield  {author} {\bibinfo {author} {\bibfnamefont {A.}~\bibnamefont {Pepper}}, \bibinfo {author} {\bibfnamefont {N.}~\bibnamefont {Tischler}},\ and\ \bibinfo {author} {\bibfnamefont {G.~J.}\ \bibnamefont {Pryde}},\ }\href {https://doi.org/10.1103/PhysRevLett.122.060501} {\bibfield  {journal} {\bibinfo  {journal} {Phys. Rev. Lett.}\ }\textbf {\bibinfo {volume} {122}},\ \bibinfo {pages} {060501} (\bibinfo {year} {2019})}\BibitemShut {NoStop}%
\bibitem [{\citenamefont {Huang}\ \emph {et~al.}(2020)\citenamefont {Huang}, \citenamefont {Ma}, \citenamefont {Yin}, \citenamefont {Tang}, \citenamefont {Dong}, \citenamefont {Chen}, \citenamefont {Xiang}, \citenamefont {Li},\ and\ \citenamefont {Guo}}]{PhysRevA.102.032412}%
  \BibitemOpen
  \bibfield  {author} {\bibinfo {author} {\bibfnamefont {C.-J.}\ \bibnamefont {Huang}}, \bibinfo {author} {\bibfnamefont {H.}~\bibnamefont {Ma}}, \bibinfo {author} {\bibfnamefont {Q.}~\bibnamefont {Yin}}, \bibinfo {author} {\bibfnamefont {J.-F.}\ \bibnamefont {Tang}}, \bibinfo {author} {\bibfnamefont {D.}~\bibnamefont {Dong}}, \bibinfo {author} {\bibfnamefont {C.}~\bibnamefont {Chen}}, \bibinfo {author} {\bibfnamefont {G.-Y.}\ \bibnamefont {Xiang}}, \bibinfo {author} {\bibfnamefont {C.-F.}\ \bibnamefont {Li}},\ and\ \bibinfo {author} {\bibfnamefont {G.-C.}\ \bibnamefont {Guo}},\ }\href {https://doi.org/10.1103/PhysRevA.102.032412} {\bibfield  {journal} {\bibinfo  {journal} {Phys. Rev. A}\ }\textbf {\bibinfo {volume} {102}},\ \bibinfo {pages} {032412} (\bibinfo {year} {2020})}\BibitemShut {NoStop}%
\bibitem [{\citenamefont {Ding}\ \emph {et~al.}(2019)\citenamefont {Ding}, \citenamefont {Lamata}, \citenamefont {Sanz}, \citenamefont {Chen},\ and\ \citenamefont {Solano}}]{https://doi.org/10.1002/qute.201800065}%
  \BibitemOpen
  \bibfield  {author} {\bibinfo {author} {\bibfnamefont {Y.}~\bibnamefont {Ding}}, \bibinfo {author} {\bibfnamefont {L.}~\bibnamefont {Lamata}}, \bibinfo {author} {\bibfnamefont {M.}~\bibnamefont {Sanz}}, \bibinfo {author} {\bibfnamefont {X.}~\bibnamefont {Chen}},\ and\ \bibinfo {author} {\bibfnamefont {E.}~\bibnamefont {Solano}},\ }\href {https://doi.org/https://doi.org/10.1002/qute.201800065} {\bibfield  {journal} {\bibinfo  {journal} {Advanced Quantum Technologies}\ }\textbf {\bibinfo {volume} {2}},\ \bibinfo {pages} {1800065} (\bibinfo {year} {2019})},\ \Eprint {https://arxiv.org/abs/https://onlinelibrary.wiley.com/doi/pdf/10.1002/qute.201800065} {https://onlinelibrary.wiley.com/doi/pdf/10.1002/qute.201800065} \BibitemShut {NoStop}%
\bibitem [{\citenamefont {Locher}\ \emph {et~al.}(2023)\citenamefont {Locher}, \citenamefont {Cardarelli},\ and\ \citenamefont {M{\"{u}}ller}}]{Locher2023quantumerror}%
  \BibitemOpen
  \bibfield  {author} {\bibinfo {author} {\bibfnamefont {D.~F.}\ \bibnamefont {Locher}}, \bibinfo {author} {\bibfnamefont {L.}~\bibnamefont {Cardarelli}},\ and\ \bibinfo {author} {\bibfnamefont {M.}~\bibnamefont {M{\"{u}}ller}},\ }\href {https://doi.org/10.22331/q-2023-03-09-942} {\bibfield  {journal} {\bibinfo  {journal} {{Quantum}}\ }\textbf {\bibinfo {volume} {7}},\ \bibinfo {pages} {942} (\bibinfo {year} {2023})}\BibitemShut {NoStop}%
\bibitem [{\citenamefont {Zhang}\ \emph {et~al.}(2021)\citenamefont {Zhang}, \citenamefont {Kong}, \citenamefont {Farooq}, \citenamefont {Yung}, \citenamefont {Guo},\ and\ \citenamefont {Wang}}]{PhysRevA.103.L040403}%
  \BibitemOpen
  \bibfield  {author} {\bibinfo {author} {\bibfnamefont {X.-M.}\ \bibnamefont {Zhang}}, \bibinfo {author} {\bibfnamefont {W.}~\bibnamefont {Kong}}, \bibinfo {author} {\bibfnamefont {M.~U.}\ \bibnamefont {Farooq}}, \bibinfo {author} {\bibfnamefont {M.-H.}\ \bibnamefont {Yung}}, \bibinfo {author} {\bibfnamefont {G.}~\bibnamefont {Guo}},\ and\ \bibinfo {author} {\bibfnamefont {X.}~\bibnamefont {Wang}},\ }\href {https://doi.org/10.1103/PhysRevA.103.L040403} {\bibfield  {journal} {\bibinfo  {journal} {Phys. Rev. A}\ }\textbf {\bibinfo {volume} {103}},\ \bibinfo {pages} {L040403} (\bibinfo {year} {2021})}\BibitemShut {NoStop}%
\bibitem [{\citenamefont {Bondarenko}\ and\ \citenamefont {Feldmann}(2020)}]{PhysRevLett.124.130502}%
  \BibitemOpen
  \bibfield  {author} {\bibinfo {author} {\bibfnamefont {D.}~\bibnamefont {Bondarenko}}\ and\ \bibinfo {author} {\bibfnamefont {P.}~\bibnamefont {Feldmann}},\ }\href {https://doi.org/10.1103/PhysRevLett.124.130502} {\bibfield  {journal} {\bibinfo  {journal} {Phys. Rev. Lett.}\ }\textbf {\bibinfo {volume} {124}},\ \bibinfo {pages} {130502} (\bibinfo {year} {2020})}\BibitemShut {NoStop}%
\bibitem [{\citenamefont {Baul}\ \emph {et~al.}(2023)\citenamefont {Baul}, \citenamefont {Walker}, \citenamefont {Moreno},\ and\ \citenamefont {Tam}}]{PhysRevE.107.045301}%
  \BibitemOpen
  \bibfield  {author} {\bibinfo {author} {\bibfnamefont {A.}~\bibnamefont {Baul}}, \bibinfo {author} {\bibfnamefont {N.}~\bibnamefont {Walker}}, \bibinfo {author} {\bibfnamefont {J.}~\bibnamefont {Moreno}},\ and\ \bibinfo {author} {\bibfnamefont {K.-M.}\ \bibnamefont {Tam}},\ }\href {https://doi.org/10.1103/PhysRevE.107.045301} {\bibfield  {journal} {\bibinfo  {journal} {Phys. Rev. E}\ }\textbf {\bibinfo {volume} {107}},\ \bibinfo {pages} {045301} (\bibinfo {year} {2023})}\BibitemShut {NoStop}%
\bibitem [{\citenamefont {Ch'ng}\ \emph {et~al.}(2018)\citenamefont {Ch'ng}, \citenamefont {Vazquez},\ and\ \citenamefont {Khatami}}]{PhysRevE.97.013306}%
  \BibitemOpen
  \bibfield  {author} {\bibinfo {author} {\bibfnamefont {K.}~\bibnamefont {Ch'ng}}, \bibinfo {author} {\bibfnamefont {N.}~\bibnamefont {Vazquez}},\ and\ \bibinfo {author} {\bibfnamefont {E.}~\bibnamefont {Khatami}},\ }\href {https://doi.org/10.1103/PhysRevE.97.013306} {\bibfield  {journal} {\bibinfo  {journal} {Phys. Rev. E}\ }\textbf {\bibinfo {volume} {97}},\ \bibinfo {pages} {013306} (\bibinfo {year} {2018})}\BibitemShut {NoStop}%
\bibitem [{\citenamefont {Eberz}\ \emph {et~al.}(2023)\citenamefont {Eberz}, \citenamefont {Link}, \citenamefont {Kell}, \citenamefont {Breyer}, \citenamefont {Gao},\ and\ \citenamefont {K\"ohl}}]{PhysRevA.108.063303}%
  \BibitemOpen
  \bibfield  {author} {\bibinfo {author} {\bibfnamefont {D.}~\bibnamefont {Eberz}}, \bibinfo {author} {\bibfnamefont {M.}~\bibnamefont {Link}}, \bibinfo {author} {\bibfnamefont {A.}~\bibnamefont {Kell}}, \bibinfo {author} {\bibfnamefont {M.}~\bibnamefont {Breyer}}, \bibinfo {author} {\bibfnamefont {K.}~\bibnamefont {Gao}},\ and\ \bibinfo {author} {\bibfnamefont {M.}~\bibnamefont {K\"ohl}},\ }\href {https://doi.org/10.1103/PhysRevA.108.063303} {\bibfield  {journal} {\bibinfo  {journal} {Phys. Rev. A}\ }\textbf {\bibinfo {volume} {108}},\ \bibinfo {pages} {063303} (\bibinfo {year} {2023})}\BibitemShut {NoStop}%
\bibitem [{\citenamefont {Kharkov}\ \emph {et~al.}(2020)\citenamefont {Kharkov}, \citenamefont {Sotskov}, \citenamefont {Karazeev}, \citenamefont {Kiktenko},\ and\ \citenamefont {Fedorov}}]{PhysRevB.101.064406}%
  \BibitemOpen
  \bibfield  {author} {\bibinfo {author} {\bibfnamefont {Y.~A.}\ \bibnamefont {Kharkov}}, \bibinfo {author} {\bibfnamefont {V.~E.}\ \bibnamefont {Sotskov}}, \bibinfo {author} {\bibfnamefont {A.~A.}\ \bibnamefont {Karazeev}}, \bibinfo {author} {\bibfnamefont {E.~O.}\ \bibnamefont {Kiktenko}},\ and\ \bibinfo {author} {\bibfnamefont {A.~K.}\ \bibnamefont {Fedorov}},\ }\href {https://doi.org/10.1103/PhysRevB.101.064406} {\bibfield  {journal} {\bibinfo  {journal} {Phys. Rev. B}\ }\textbf {\bibinfo {volume} {101}},\ \bibinfo {pages} {064406} (\bibinfo {year} {2020})}\BibitemShut {NoStop}%
\bibitem [{\citenamefont {Wu}\ \emph {et~al.}(2021)\citenamefont {Wu}, \citenamefont {Zhang},\ and\ \citenamefont {Zhai}}]{PhysRevResearch.3.L032057}%
  \BibitemOpen
  \bibfield  {author} {\bibinfo {author} {\bibfnamefont {Y.}~\bibnamefont {Wu}}, \bibinfo {author} {\bibfnamefont {P.}~\bibnamefont {Zhang}},\ and\ \bibinfo {author} {\bibfnamefont {H.}~\bibnamefont {Zhai}},\ }\href {https://doi.org/10.1103/PhysRevResearch.3.L032057} {\bibfield  {journal} {\bibinfo  {journal} {Phys. Rev. Res.}\ }\textbf {\bibinfo {volume} {3}},\ \bibinfo {pages} {L032057} (\bibinfo {year} {2021})}\BibitemShut {NoStop}%
\bibitem [{\citenamefont {Ghosh}\ and\ \citenamefont {Ghosh}(2023)}]{PhysRevB.108.165408}%
  \BibitemOpen
  \bibfield  {author} {\bibinfo {author} {\bibfnamefont {K.~J.~B.}\ \bibnamefont {Ghosh}}\ and\ \bibinfo {author} {\bibfnamefont {S.}~\bibnamefont {Ghosh}},\ }\href {https://doi.org/10.1103/PhysRevB.108.165408} {\bibfield  {journal} {\bibinfo  {journal} {Phys. Rev. B}\ }\textbf {\bibinfo {volume} {108}},\ \bibinfo {pages} {165408} (\bibinfo {year} {2023})}\BibitemShut {NoStop}%
\bibitem [{\citenamefont {Kottmann}\ \emph {et~al.}(2021)\citenamefont {Kottmann}, \citenamefont {Metz}, \citenamefont {Fraxanet},\ and\ \citenamefont {Baldelli}}]{PhysRevResearch.3.043184}%
  \BibitemOpen
  \bibfield  {author} {\bibinfo {author} {\bibfnamefont {K.}~\bibnamefont {Kottmann}}, \bibinfo {author} {\bibfnamefont {F.}~\bibnamefont {Metz}}, \bibinfo {author} {\bibfnamefont {J.}~\bibnamefont {Fraxanet}},\ and\ \bibinfo {author} {\bibfnamefont {N.}~\bibnamefont {Baldelli}},\ }\href {https://doi.org/10.1103/PhysRevResearch.3.043184} {\bibfield  {journal} {\bibinfo  {journal} {Phys. Rev. Res.}\ }\textbf {\bibinfo {volume} {3}},\ \bibinfo {pages} {043184} (\bibinfo {year} {2021})}\BibitemShut {NoStop}%
\bibitem [{\citenamefont {Ngairangbam}\ \emph {et~al.}(2022)\citenamefont {Ngairangbam}, \citenamefont {Spannowsky},\ and\ \citenamefont {Takeuchi}}]{PhysRevD.105.095004}%
  \BibitemOpen
  \bibfield  {author} {\bibinfo {author} {\bibfnamefont {V.~S.}\ \bibnamefont {Ngairangbam}}, \bibinfo {author} {\bibfnamefont {M.}~\bibnamefont {Spannowsky}},\ and\ \bibinfo {author} {\bibfnamefont {M.}~\bibnamefont {Takeuchi}},\ }\href {https://doi.org/10.1103/PhysRevD.105.095004} {\bibfield  {journal} {\bibinfo  {journal} {Phys. Rev. D}\ }\textbf {\bibinfo {volume} {105}},\ \bibinfo {pages} {095004} (\bibinfo {year} {2022})}\BibitemShut {NoStop}%
\bibitem [{\citenamefont {Luchi}\ \emph {et~al.}(2023)\citenamefont {Luchi}, \citenamefont {Trevisanutto}, \citenamefont {Roggero}, \citenamefont {DuBois}, \citenamefont {Rosen}, \citenamefont {Turro}, \citenamefont {Amitrano},\ and\ \citenamefont {Pederiva}}]{PhysRevApplied.20.014045}%
  \BibitemOpen
  \bibfield  {author} {\bibinfo {author} {\bibfnamefont {P.}~\bibnamefont {Luchi}}, \bibinfo {author} {\bibfnamefont {P.~E.}\ \bibnamefont {Trevisanutto}}, \bibinfo {author} {\bibfnamefont {A.}~\bibnamefont {Roggero}}, \bibinfo {author} {\bibfnamefont {J.~L.}\ \bibnamefont {DuBois}}, \bibinfo {author} {\bibfnamefont {Y.~J.}\ \bibnamefont {Rosen}}, \bibinfo {author} {\bibfnamefont {F.}~\bibnamefont {Turro}}, \bibinfo {author} {\bibfnamefont {V.}~\bibnamefont {Amitrano}},\ and\ \bibinfo {author} {\bibfnamefont {F.}~\bibnamefont {Pederiva}},\ }\href {https://doi.org/10.1103/PhysRevApplied.20.014045} {\bibfield  {journal} {\bibinfo  {journal} {Phys. Rev. Appl.}\ }\textbf {\bibinfo {volume} {20}},\ \bibinfo {pages} {014045} (\bibinfo {year} {2023})}\BibitemShut {NoStop}%
\bibitem [{\citenamefont {Schmitt}\ and\ \citenamefont {Lenar\ifmmode \check{c}\else \v{c}\fi{}i\ifmmode~\check{c}\else \v{c}\fi{}}(2022)}]{PhysRevB.106.L041110}%
  \BibitemOpen
  \bibfield  {author} {\bibinfo {author} {\bibfnamefont {M.}~\bibnamefont {Schmitt}}\ and\ \bibinfo {author} {\bibfnamefont {Z.}~\bibnamefont {Lenar\ifmmode \check{c}\else \v{c}\fi{}i\ifmmode~\check{c}\else \v{c}\fi{}}},\ }\href {https://doi.org/10.1103/PhysRevB.106.L041110} {\bibfield  {journal} {\bibinfo  {journal} {Phys. Rev. B}\ }\textbf {\bibinfo {volume} {106}},\ \bibinfo {pages} {L041110} (\bibinfo {year} {2022})}\BibitemShut {NoStop}%
\bibitem [{\citenamefont {Szo\l{}dra}\ \emph {et~al.}(2022)\citenamefont {Szo\l{}dra}, \citenamefont {Sierant}, \citenamefont {Lewenstein},\ and\ \citenamefont {Zakrzewski}}]{PhysRevB.105.224205}%
  \BibitemOpen
  \bibfield  {author} {\bibinfo {author} {\bibfnamefont {T.}~\bibnamefont {Szo\l{}dra}}, \bibinfo {author} {\bibfnamefont {P.}~\bibnamefont {Sierant}}, \bibinfo {author} {\bibfnamefont {M.}~\bibnamefont {Lewenstein}},\ and\ \bibinfo {author} {\bibfnamefont {J.}~\bibnamefont {Zakrzewski}},\ }\href {https://doi.org/10.1103/PhysRevB.105.224205} {\bibfield  {journal} {\bibinfo  {journal} {Phys. Rev. B}\ }\textbf {\bibinfo {volume} {105}},\ \bibinfo {pages} {224205} (\bibinfo {year} {2022})}\BibitemShut {NoStop}%
\bibitem [{\citenamefont {Sone}\ \emph {et~al.}(2023)\citenamefont {Sone}, \citenamefont {Yamamoto}, \citenamefont {Holdsworth},\ and\ \citenamefont {Narang}}]{PhysRevResearch.5.023039}%
  \BibitemOpen
  \bibfield  {author} {\bibinfo {author} {\bibfnamefont {A.}~\bibnamefont {Sone}}, \bibinfo {author} {\bibfnamefont {N.}~\bibnamefont {Yamamoto}}, \bibinfo {author} {\bibfnamefont {T.}~\bibnamefont {Holdsworth}},\ and\ \bibinfo {author} {\bibfnamefont {P.}~\bibnamefont {Narang}},\ }\href {https://doi.org/10.1103/PhysRevResearch.5.023039} {\bibfield  {journal} {\bibinfo  {journal} {Phys. Rev. Res.}\ }\textbf {\bibinfo {volume} {5}},\ \bibinfo {pages} {023039} (\bibinfo {year} {2023})}\BibitemShut {NoStop}%
\bibitem [{\citenamefont {Patel}\ \emph {et~al.}(2023)\citenamefont {Patel}, \citenamefont {Collis}, \citenamefont {Duong}, \citenamefont {Koch}, \citenamefont {Cutugno}, \citenamefont {Wessing},\ and\ \citenamefont {Alsing}}]{Patel_2023}%
  \BibitemOpen
  \bibfield  {author} {\bibinfo {author} {\bibfnamefont {S.}~\bibnamefont {Patel}}, \bibinfo {author} {\bibfnamefont {B.}~\bibnamefont {Collis}}, \bibinfo {author} {\bibfnamefont {W.}~\bibnamefont {Duong}}, \bibinfo {author} {\bibfnamefont {D.}~\bibnamefont {Koch}}, \bibinfo {author} {\bibfnamefont {M.}~\bibnamefont {Cutugno}}, \bibinfo {author} {\bibfnamefont {L.}~\bibnamefont {Wessing}},\ and\ \bibinfo {author} {\bibfnamefont {P.}~\bibnamefont {Alsing}},\ }\href {https://doi.org/10.1088/1402-4896/acc492} {\bibfield  {journal} {\bibinfo  {journal} {Physica Scripta}\ }\textbf {\bibinfo {volume} {98}},\ \bibinfo {pages} {045111} (\bibinfo {year} {2023})}\BibitemShut {NoStop}%
\bibitem [{\citenamefont {Cao}\ and\ \citenamefont {Wang}(2021)}]{PhysRevApplied.15.054012}%
  \BibitemOpen
  \bibfield  {author} {\bibinfo {author} {\bibfnamefont {C.}~\bibnamefont {Cao}}\ and\ \bibinfo {author} {\bibfnamefont {X.}~\bibnamefont {Wang}},\ }\href {https://doi.org/10.1103/PhysRevApplied.15.054012} {\bibfield  {journal} {\bibinfo  {journal} {Phys. Rev. Appl.}\ }\textbf {\bibinfo {volume} {15}},\ \bibinfo {pages} {054012} (\bibinfo {year} {2021})}\BibitemShut {NoStop}%
\bibitem [{\citenamefont {Hiai}\ and\ \citenamefont {Petz}(1991)}]{hiai1991proper}%
  \BibitemOpen
  \bibfield  {author} {\bibinfo {author} {\bibfnamefont {F.}~\bibnamefont {Hiai}}\ and\ \bibinfo {author} {\bibfnamefont {D.}~\bibnamefont {Petz}},\ }\href@noop {} {\bibfield  {journal} {\bibinfo  {journal} {Communications in mathematical physics}\ }\textbf {\bibinfo {volume} {143}},\ \bibinfo {pages} {99} (\bibinfo {year} {1991})}\BibitemShut {NoStop}%
\bibitem [{\citenamefont {Lesniewski}\ and\ \citenamefont {Ruskai}(1999)}]{10.1063/1.533053}%
  \BibitemOpen
  \bibfield  {author} {\bibinfo {author} {\bibfnamefont {A.}~\bibnamefont {Lesniewski}}\ and\ \bibinfo {author} {\bibfnamefont {M.~B.}\ \bibnamefont {Ruskai}},\ }\href {https://doi.org/10.1063/1.533053} {\bibfield  {journal} {\bibinfo  {journal} {Journal of Mathematical Physics}\ }\textbf {\bibinfo {volume} {40}},\ \bibinfo {pages} {5702} (\bibinfo {year} {1999})},\ \Eprint {https://arxiv.org/abs/https://pubs.aip.org/aip/jmp/article-pdf/40/11/5702/19013936/5702\_1\_online.pdf} {https://pubs.aip.org/aip/jmp/article-pdf/40/11/5702/19013936/5702\_1\_online.pdf} \BibitemShut {NoStop}%
\bibitem [{\citenamefont {Berta}\ and\ \citenamefont {Majenz}(2018)}]{PhysRevLett.121.190503}%
  \BibitemOpen
  \bibfield  {author} {\bibinfo {author} {\bibfnamefont {M.}~\bibnamefont {Berta}}\ and\ \bibinfo {author} {\bibfnamefont {C.}~\bibnamefont {Majenz}},\ }\href {https://doi.org/10.1103/PhysRevLett.121.190503} {\bibfield  {journal} {\bibinfo  {journal} {Phys. Rev. Lett.}\ }\textbf {\bibinfo {volume} {121}},\ \bibinfo {pages} {190503} (\bibinfo {year} {2018})}\BibitemShut {NoStop}%
\bibitem [{\citenamefont {Anshu}\ \emph {et~al.}(2018)\citenamefont {Anshu}, \citenamefont {Hsieh},\ and\ \citenamefont {Jain}}]{PhysRevLett.121.190504}%
  \BibitemOpen
  \bibfield  {author} {\bibinfo {author} {\bibfnamefont {A.}~\bibnamefont {Anshu}}, \bibinfo {author} {\bibfnamefont {M.-H.}\ \bibnamefont {Hsieh}},\ and\ \bibinfo {author} {\bibfnamefont {R.}~\bibnamefont {Jain}},\ }\href {https://doi.org/10.1103/PhysRevLett.121.190504} {\bibfield  {journal} {\bibinfo  {journal} {Phys. Rev. Lett.}\ }\textbf {\bibinfo {volume} {121}},\ \bibinfo {pages} {190504} (\bibinfo {year} {2018})}\BibitemShut {NoStop}%
\bibitem [{\citenamefont {Salazar}(2024)}]{PhysRevE.109.L012103}%
  \BibitemOpen
  \bibfield  {author} {\bibinfo {author} {\bibfnamefont {D.~S.~P.}\ \bibnamefont {Salazar}},\ }\href {https://doi.org/10.1103/PhysRevE.109.L012103} {\bibfield  {journal} {\bibinfo  {journal} {Phys. Rev. E}\ }\textbf {\bibinfo {volume} {109}},\ \bibinfo {pages} {L012103} (\bibinfo {year} {2024})}\BibitemShut {NoStop}%
\bibitem [{\citenamefont {Mu}\ and\ \citenamefont {Li}(2020)}]{PhysRevA.102.022217}%
  \BibitemOpen
  \bibfield  {author} {\bibinfo {author} {\bibfnamefont {H.}~\bibnamefont {Mu}}\ and\ \bibinfo {author} {\bibfnamefont {Y.}~\bibnamefont {Li}},\ }\href {https://doi.org/10.1103/PhysRevA.102.022217} {\bibfield  {journal} {\bibinfo  {journal} {Phys. Rev. A}\ }\textbf {\bibinfo {volume} {102}},\ \bibinfo {pages} {022217} (\bibinfo {year} {2020})}\BibitemShut {NoStop}%
\bibitem [{\citenamefont {Genoni}\ \emph {et~al.}(2008)\citenamefont {Genoni}, \citenamefont {Paris},\ and\ \citenamefont {Banaszek}}]{PhysRevA.78.060303}%
  \BibitemOpen
  \bibfield  {author} {\bibinfo {author} {\bibfnamefont {M.~G.}\ \bibnamefont {Genoni}}, \bibinfo {author} {\bibfnamefont {M.~G.~A.}\ \bibnamefont {Paris}},\ and\ \bibinfo {author} {\bibfnamefont {K.}~\bibnamefont {Banaszek}},\ }\href {https://doi.org/10.1103/PhysRevA.78.060303} {\bibfield  {journal} {\bibinfo  {journal} {Phys. Rev. A}\ }\textbf {\bibinfo {volume} {78}},\ \bibinfo {pages} {060303} (\bibinfo {year} {2008})}\BibitemShut {NoStop}%
\bibitem [{\citenamefont {Yang}\ \emph {et~al.}(2023)\citenamefont {Yang}, \citenamefont {Xu},\ and\ \citenamefont {Zhou}}]{PhysRevA.108.052402}%
  \BibitemOpen
  \bibfield  {author} {\bibinfo {author} {\bibfnamefont {M.}~\bibnamefont {Yang}}, \bibinfo {author} {\bibfnamefont {C.-Q.}\ \bibnamefont {Xu}},\ and\ \bibinfo {author} {\bibfnamefont {D.~L.}\ \bibnamefont {Zhou}},\ }\href {https://doi.org/10.1103/PhysRevA.108.052402} {\bibfield  {journal} {\bibinfo  {journal} {Phys. Rev. A}\ }\textbf {\bibinfo {volume} {108}},\ \bibinfo {pages} {052402} (\bibinfo {year} {2023})}\BibitemShut {NoStop}%
\bibitem [{\citenamefont {Bhatia}(2012)}]{matrix}%
  \BibitemOpen
  \bibfield  {author} {\bibinfo {author} {\bibfnamefont {R.}~\bibnamefont {Bhatia}},\ }\href@noop {} {\emph {\bibinfo {title} {Matrix Analysis}}}\ (\bibinfo  {publisher} {Springer New York, NY},\ \bibinfo {year} {2012})\BibitemShut {NoStop}%
\bibitem [{\citenamefont {Andrews}(1984)}]{andrews_1984}%
  \BibitemOpen
  \bibfield  {author} {\bibinfo {author} {\bibfnamefont {G.~E.}\ \bibnamefont {Andrews}},\ }\href {https://doi.org/10.1017/CBO9780511608650} {\emph {\bibinfo {title} {The Theory of Partitions}}},\ Encyclopedia of Mathematics and its Applications\ (\bibinfo  {publisher} {Cambridge University Press},\ \bibinfo {year} {1984})\BibitemShut {NoStop}%
\bibitem [{SM()}]{SM}%
  \BibitemOpen
  \href@noop {} {}\bibinfo {note} {See Supplemental Material}\BibitemShut {NoStop}%
\end{thebibliography}%
\widetext
\clearpage
\begin{center}
\textbf{\large \centering Supplemental Material to ``Optimized Quantum Autoencoder''}
\end{center}

\vspace{0.3cm}
In this supplementary material, we provide the algorithm used in the paper ``Optimized Quantum Autoencoder''.

\section{Regular Young tableaux generation algorithm}
Because the Young tableau we are considering is a table with $d_{A}$ rows and $d_{B}$ columns, with entries from 1 to $d_{A}d_{B}$, such that each row has elements increasing from left to right, and each column has elements increasing from top to bottom. We can generate a Young tableau by sequentially filling the table with numbers from 1 to $d_{A}d_{B}$. However, due to the incremental nature, when we sequentially fill the table with numbers from 1 to $d_{A}d_{B}$, if there are cells above or to the left of the cell we are about to fill, those cells must already be filled. Using this property, we can obtain the following Algorithm \ref{alg_1}. 
\begin{algorithm}
\label{alg_1}
    \caption{Regular Young tableaux generation algorithm}
    \KwIn{The dimension of system A and system B: $d_{A},d_{B}$}
    \KwOut{$MatrixSet$ containing all $d_{A}\times d_{B}$ regular Young tableaux}
    $S=d_{A}\times d_{B}$ matrix with all its elements being 0\;
    \If{$d_{A}=d_{B}$}{$S[1][1] \gets 1$\;
    $S[1][2] \gets 2$\;
    $MatrixSet \leftarrow \{ S \}$\;
    \For{$i_{n} \gets 3$ \KwTo $d_{A}\times d_{B}$}
    {$MatrixSet_{T}={\rm empty~set}$\;
    \For{$matrix \in MatrixSet$}
    {\For{$i \gets 1$ \KwTo $d_{A}$}
    {\For{$j \gets 1$ \KwTo $d_{B}$}
    {\If{$matrix[i-1][j]\  and\  matrix[i][j-1]$ are empty or not zero}{$matrix[i][j]\gets i_{n}$\;
    $MatrixSet_{T} \leftarrow MatrixSet_{T} \cup matrix$}}
    }    
    }    
    $MatrixSet\gets MatrixSet_{T}$
    }    
    }    
    \Else{$S[1][1] \gets 1$\;
    $MatrixSet \leftarrow \{ S \}$
    \For{$i_{n} \gets 2$ \KwTo $d_{A}\times d_{B}$}
    {$MatrixSet_{T}={\rm empty~set}$\;
    \For{$matrix \in MatrixSet$}
    {\For{$i \gets 1$ \KwTo $d_{A}$}
    {\For{$j \gets 1$ \KwTo $d_{B}$}
    {\If{$matrix[i-1][j]\  and\  matrix[i][j-1]$ are empty or not zero}{$matrix[i][j]\gets i_{n}$\;
    $MatrixSet_{T} \leftarrow MatrixSet_{T} \cup matrix$}}
    }   
    }
    $MatrixSet\gets MatrixSet_{T}$
    }
    }
    \Return $MatrixSet$
\end{algorithm}
The reason we divide it into two cases is that when $d_{A}$ equals $d_{B}$, due to the symmetry of the AB system, we only need to consider half of the regular Young tableaux.

\section{Search Algorithm}
Algorithm \ref{alg_1} will fail when the number of regular Young tableaux exceeds the range of our computational capacity. Fortunately, we can use breadth-first search algorithms and depth-first search algorithms to solve this problem. Although this algorithm may not necessarily find the minimum value of mutual information, it is relatively superior to the variational quantum circuit method.

\begin{figure}
\vspace{3mm}
\centering
\includegraphics[width=0.45\textwidth]{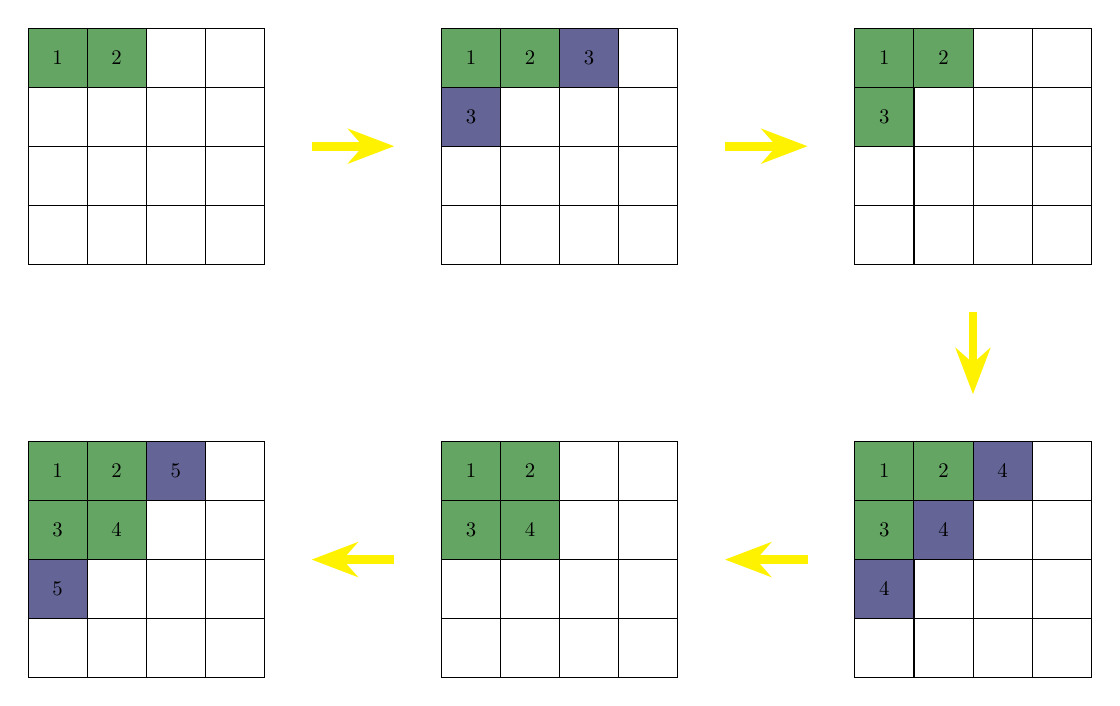}
\caption{
Taking a $4\times 4$ Young tableau as an example, we illustrate the process of randomly generating a regular Young tableau. Since a regular Young tableau satisfies the condition of increasing sequentially from top to bottom and from left to right, when we sequentially fill in the numbers 1 to 16 into the tableau, it is imperative to ensure that before each insertion, the left and upper cells of the target cell are already filled with numbers. Following this rule, we can determine which cells are available for filling in each step, marked in purple. Finally, a random selection is made from these purple cells to insert a number, resulting in the random generation of a regular Young tableau.
}
\label{2_a}
\end{figure}

The breadth-first search algorithm is based on the construction of regular Young tableaux. As shown in Fig.~\ref{2_a}, we fill the numbers 1 to $d_{A}d_{B}$ into the table. Each time we fill a number, for the available positions for that number, we randomly choose a position with uniform probability to fill the number. In the end, we will randomly generate a regular Young tableau. In the breadth-first search algorithm, following the aforementioned method, we randomly generate $N_{1}$ regular Young tableaux while simultaneously calculating the mutual information corresponding to these $N_{1}$ tableaux. Finally, we select $N_{2}$ regular Young tableaux with the minimum mutual information among them. The specific implementation of the algorithm is provided in Algorithm \ref{alg_2}.
\begin{figure}
\vspace{3mm}
\centering
\includegraphics[width=0.45\textwidth]{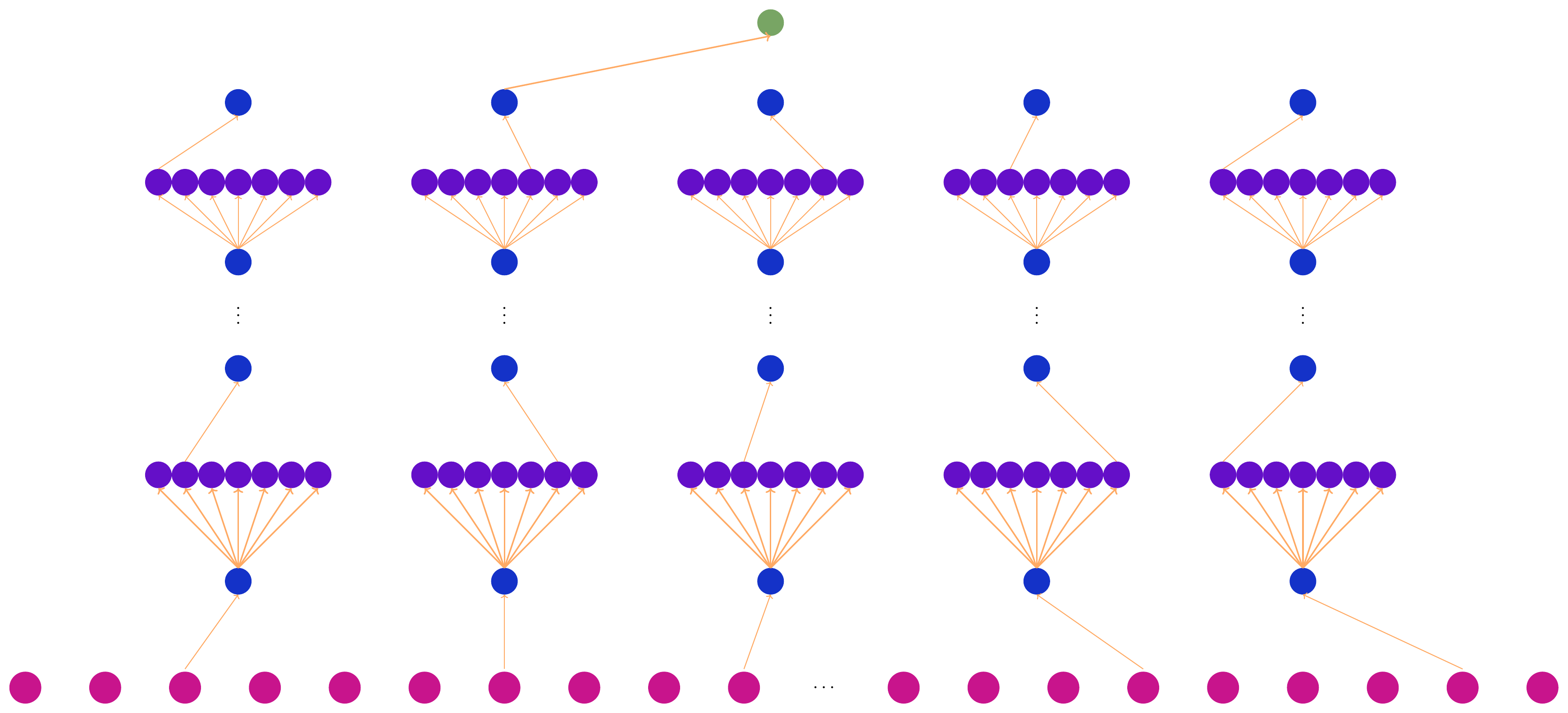}
\caption{
Our search algorithm is depicted in the figure. Initially, we randomly generate $N_1$ regular Young tableaux using Algorithm \ref{alg_2}, represented by the red sphere. Subsequently, We select $N_2$ regular Young tableaux with the minimum corresponding mutual information from the set of $N_1$ regular Young tableaux, indicated by the blue spheres at the bottom of the figure which only displays five spheres. Following this, starting from the blue tableaux, we swap the numbers i with i+1 and i+2 within each tableau. If the result remains a regular Young tableau, we record it in the purple tableaux. Finally, from the various groups of purple tableaux, we select the tableau with the minimum mutual information to obtain the next layer of blue tableaux. This process is repeated $N_D$ times, and from the final set of blue regular Young tableaux, we choose the tableau with the lowest corresponding mutual information as our optimized result.
}
\label{2_b}
\end{figure}

\begin{algorithm}
\label{alg_2}
    \caption{Breadth-first search algorithm}
    \KwIn{$d_A d_B$ eigenvalues of initial state, number of searches $N_1$ and number of seeds reserved for depth-first search $N_2$}
    \KwOut{$N_2$ regular Young tableaux with lower corresponding mutual information}
    $S=d_{A}\times d_{B}$ matrix with all its elements being $0$\;
    $S[1][1] \gets 1$\;
    $MatrixSet={\rm empty~set}$\;
    \For{$n \gets 1$ \KwTo $N_1$}
    {$M \leftarrow S$\;
    \For{$i_s \gets 2$ \KwTo $d_A d_B$ }
    {$matrixset={\rm empty~set}$\;
    \For{$i \gets 1$ \KwTo $d_A$}
    {\For{$j \gets 1$ \KwTo $d_B$}
    {\If{$M[i-1][j]$~\rm and~$M[i][j-1]$ \rm are empty or not zero}{$S_T\leftarrow M$\;
    $S_T[i][j]\leftarrow i_s$\;
    $matrixset \leftarrow matrixset \cup S_T$}}
    }
    $M\leftarrow$ Randomly select a matrix from $matrixset$
    }
    $MatrixSet \leftarrow MatrixSet \cup M$
    }
    \Return The $N_2$ regular Young tableaux with lower corresponding mutual information in $MatrixSet$   
\end{algorithm}

In the breadth-first search algorithm, we obtain $N_{2}$ regular Young tableaux, and the mutual information corresponding to these $N_{2}$ tableaux is already relatively low. The depth-first search algorithm involves searching around these $N_{2}$ regular Young tableaux to find regular Young tableaux with even lower mutual information.

We use this method to generate regular Young tableaux around a given regular Young tableau $Y$. For a given tableau $Y$, we locate cells filled with numbers $i,~i+1$ and $i+2$. We then create a tableau $Y^{1}_{i}$ by exchanging the numbers filled in the cells for $i$ and $i+1$, and a tableau $Y^{2}_{i}$ by exchanging the numbers filled in the cells for $i$ and $i+2$. We iterate over $i$ to obtain a set of tableaux, and then retain only the regular Young tableaux from this set. These tableaux constitute the regular Young tableaux around the given tableau. As depicted in Fig.~\ref{2_b}, in the depth-first search algorithm, we only retain the regular Young tableau with the lowest mutual information from these sets. Furthermore, for the retained regular Young tableaux, we repeat the above algorithm to further decrease the mutual information. The specific implementation of the algorithm is provided in Algorithm \ref{alg_3}.
\begin{algorithm}
\label{alg_3}
    \caption{Depth-first search algorithm}
    \KwIn{The $N_2$ regular Young tableaux obtained using breadth-first search algorithm and the search depth $N_D$}
    \KwOut{A regular Young tableau with lower corresponding mutual information along with its corresponding mutual information}
    $MatrixSet = N_2$ regular Young tableaux\; 
    \For{$matrix_{a}\in MatrixSet$}
    {\For{$d \gets 1$ \KwTo $N_D$}
    {$matrixset={\rm empty~set}$\;
    \For{$i \gets 2$ \KwTo $d_A d_B-1$}{
    $matrix_c \leftarrow$ Change the position of elements in the $matrix_{a}$ whose values are $i$ and $i+1$\;
    \If{$matrix_c$ \rm is a regular Young tableau}{$matrixset\leftarrow matrixset \cup matrix_c$}
    }
    \For{$i \gets 2$ \KwTo $d_A d_B-2$}{
    $matrix_c \leftarrow $Change the position of elements in the $matrix_{a}$ whose values are $i$ and $i+2$\;
    \If{$matrix_c$ is a regular Young tableau}{$matrixset\leftarrow matrixset \cup matrix_c$}
    }
    $matrix_a \leftarrow $the matrix in $matrixset$ with the lowest corresponding mutual information
    }
    }
    $matrix_{end}\leftarrow$ a regular Young tableau in $MatrixSet$ with lower corresponding mutual information\;
    \Return  $matrix_{end}$ along with its corresponding mutual information
\end{algorithm}
\section{VQC optimization}
\label{vqc:structure}
We use two types of variational quantum circuits to optimize the mutual information of quantum states. In circuit VQC\_1, each R gate is $R_{Z}R_{Y}R_{Z}$ gates. We used a total of 42 R gates, so circuit VQC\_1 has 126 parameters that can be optimized. In circuit VQC\_2, each layer has a total of 18 rotation gates, and we used 8 layers, so there are 144 parameters that can be optimized. For VQC optimization, our approach is as follows: for each quantum state, we optimize it using each circuit three times, and the final result is the best outcome among the three optimizations.

\begin{figure}
\subfigure[]{
\includegraphics[width=1\textwidth]{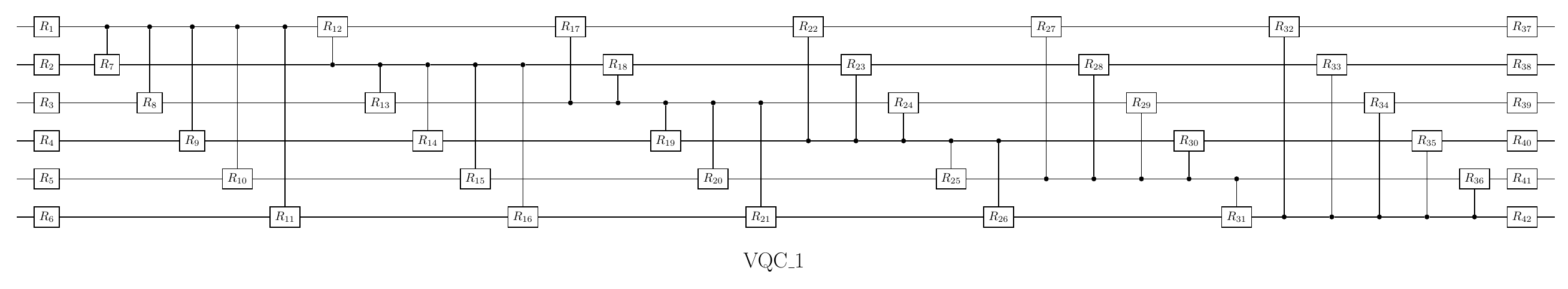}}
\label{2_d_1}
\subfigure[]{
\includegraphics[width=0.2\textwidth]{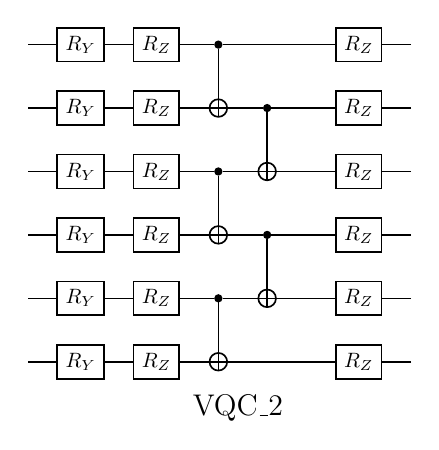}}
\label{2_d_2}
\caption{
The figure illustrates two types of variational quantum circuits. 
In the first diagram, each R gate is $R_{Z}R_{Y}R_{Z}$ gates. In the second circuit, we employed a total of 8 layers.
}
\label{2_d}
\end{figure}

\end{document}